\documentclass[%
 aip,
 jcp,
 amsmath,
 amssymb,
preprint,%
]{revtex4-1}
\usepackage[utf8]{inputenc}
\usepackage[T1]{fontenc}
\usepackage{mathptmx}
\usepackage{graphicx}
\usepackage{dcolumn}
\usepackage{bm}
\usepackage{color}
\usepackage{siunitx}

\renewcommand{\v}[1]{\ensuremath{\mathbf{#1}}} 
\newcommand{\gv}[1]{\ensuremath{\mbox{\boldmath$ #1 $}}} 

\newcommand{\figref}[1]{Fig.~\ref{#1}}
\newcommand{\siref}[1]{SI~\ref{#1}}
\newcommand{\name}[1]{_{\text{#1}}} 
\newcommand{\pd}[2]{\frac{\partial #1}{\partial #2}}

\begin{document}
\title{Hybrid Particle-Field Molecular Dynamics Under Constant Pressure}
\author{Sigbj\o rn L\o land Bore}
\email[email: ]{s.l.bore@kjemi.uio.no}
\affiliation{Department of Chemistry, and Hylleraas Centre for Quantum Molecular Sciences, University of Oslo, PO Box 1033 Blindern, 0315 Oslo, Norway} 
\author{Hima Bindu Kolli}
\altaffiliation[Present address: ]{Department of Physics and Astronomy, The University of Sheffield, United Kingdom}
\affiliation{Department of Chemistry, and Hylleraas Centre for Quantum Molecular Sciences, University of Oslo, PO Box 1033 Blindern, 0315 Oslo, Norway} 
\author{Antonio {De~Nicola}}
\affiliation{Department of Organic Materials Science, Yamagata University, 4-3-16 Jonan Yonezawa, Yamagata-ken 992-8510, Japan}
\author{Maksym Byshkin}
\affiliation{Institute of Computational Science,  Università della Svizzera italiana, 6900 Lugano, Switzerland}
\author{Toshihiro Kawakatsu}
\affiliation{Department of Physics, Tohoku University, Aoba, Aramaki, Aoba-ku,
Sendai 980-8578, Japan}
\author{Giuseppe Milano}
\affiliation{Department of Organic Materials Science, Yamagata University, 4-3-16 Jonan Yonezawa, Yamagata-ken 992-8510, Japan}
\author{Michele Cascella}
\email[email: ]{michele.cascella@kjemi.uio.no}
\affiliation{Department of Chemistry, and Hylleraas Centre for Quantum Molecular Sciences, University of Oslo, PO Box 1033 Blindern, 0315 Oslo, Norway} 
 \begin{abstract}
{\small Hybrid particle-field methods are computationally efficient approaches for modelling soft matter systems. So far applications of these methodologies have been limited to constant volume conditions. Here, we reformulate  particle-field interactions to represent systems coupled to constant external pressure.  First, we show that the commonly used particle-field energy functional can be modified to model and parameterize the isotropic contributions to the pressure tensor without interfering with the microscopic forces on the particles. Second, we employ a square gradient particle-field interaction term to model non-isotropic contributions to the pressure tensor, such as in surface tension phenomena. This formulation is implemented within the hybrid particle-field molecular dynamics approach and is tested on a series of model systems. Simulations of a homogeneous water box demonstrate that it is possible to parameterize the equation of state to reproduce any target density for a given external pressure. Moreover, the same parameterization is transferable to systems of similar coarse-grained mapping resolution. Finally, we evaluate the feasibility of the proposed approach on coarse-grained models of phospholipids, finding that the term between water and the lipid hydrocarbon tails is alone sufficient to reproduce the experimental area per lipid in constant-pressure simulations, and to produce a qualitatively correct lateral pressure profile.}
 \end{abstract}
\date{\today}
\maketitle
\section{Introduction}
Hybrid particle-field simulations (hPF) are a group of
computationally efficient approaches for studying mesoscale soft
matter systems with molecular
resolution.~\cite{Daoulas2006JCP,Muller2011JSP,Milano2009JCP,Vogiatzis2017MACRO}
In hPF models, computationally expensive, intermolecular pair
interaction potentials are replaced by an inhomogeneous external
potential that is functionally dependent on the densities of the particles
composing the system. As a consequence, the motion of the moieties
composing the system decouples, yielding a substantial simplification
for the sampling of the phase space. From an algorithmic point of
view, the hPF methods are efficiently represented by particle-mesh
approaches, giving excellent parallelization
efficiency~\cite{Zhao2012JCP}. Very recently, a GPU-based
implementation of the Monte Carlo based hPF
(\emph{single chain in mean field}) set a new milestone with simulations
of polymer melts with 10 billion particles~\cite{Schneider2019CPC}.

Coupling hPF to molecular dynamics algorithms has widened the range of
applicability of hPF systems, from more conventional soft polymer
mixtures to biological
systems~\cite{Milano2013PHYSBIO,Soares2017JPCL,Cascella2015CHEMMOD,Marrink2019CHEMREV}. Examples
from the literature include nanocomposites, nanoparticles, percolation
phenomena in carbon
nanotubes~\cite{Nicola2016EPJ,Zhao2016NANOSCALE,Munao2018NANOSCALE,Munao2019MACRO},
lamellar and nonlamellar phases of
phospholipids~\cite{Nicolia2012TCA,Nicolia2011JCTC}, and more recently
polypeptides, and
polyelectrolytes~\cite{Zhu2016PCCP,kolli2018JCTC,Bore2019JCTC,Denicola2020BBA}.

Despite the growing level of maturity reached by hPF simulations, to
the best of the authors knowledge, all works that
have appeared in the literature so far have been restricted to
canonical, constant volume ($NVT$) thermodynamic conditions. In fact, the study of many important phenomena requires targeting constant pressure
conditions ($NPT$). For example, structural and dynamic properties of
lipid membranes are typically defined at fixed tension (prominently, at zero tension), which are best represented within the
$NPT$ ensemble. Furthermore, the average density of heterogeneous or multiphase systems often cannot
straightforwardly be determined from the bulk values of its constituents,
making it difficult to establish physically sound $NVT$ conditions
in the absence of a preliminary equilibration at $NPT$, or 
of additional information from other experimental 
or computational sources. 

The main issue related to the calculation of the pressure in hPF
resides in determining the contribution by the particle-field
interaction energy. In particular, contrary to ordinary pair
potentials, such term cannot be computed from the virial of the
intermolecular forces. In 2003, Tyler and Morse~\cite{Tyler2003MACRO}
proposed a derivation of the pressure in a \emph{continuum
  self-consistent field theory} formalism by computing the change in
free energy upon a change in the volume. More recently, some of us
proposed a first formulation for pressure in hPF~\cite{Milano2010JCP}
by a virtual displacement approach~\cite{Brown1995MP}, obtaining a
good correspondence of the equation of state for polymer chains
compared to that derived from particle-based simulations. In a very
recent publication~\cite{Ting2017JCP}, Ting and M{\" u}ller also
considered local pressure profiles in multiphase systems within
self-consistent field theory, putting particular emphasis on bilayer
structures. With the added novelty of using Kirkwood-Irving assignment
of pressure contributions from bonded terms, they obtained excellent
agreement between interface properties computed from local pressure
profiles and thermodynamic considerations, demonstrating also the
usefulness of \emph{local pressure profiles} in density field based
methods.~\cite{Ting2017JCP} Finally, Sgouros et
al.~\cite{Sgouros2018Macro} derived the pressure for hPF using the
thermodynamic definition of the pressure tensor~\cite{Lustig1996JR}.

Despite the capability of deriving and computing the pressure in $NVT$
conditions, two important issues hinder hPF simulations under constant
pressure.  First, the interaction energy functionals commonly used in
hPF simulations~\cite{Milano2009JCP,Daoulas2006JCP} are not designed to give a realistic representation of
the equation of state. Second, as can be seen from inspection of
density field contributions in refs.~\cite{Ting2017JCP,Milano2010JCP}
and is emphasized in ref.~\cite{Sgouros2018Macro}, pure
density terms \emph{contribute only isotropically} to the
pressure. This is particularly detrimental for interfacial phenomena,
where the appearance of any surface tension is only limited to the
eventual non-isotropic orientation of the bonded terms for spatially
organized molecules.

In density field approaches, the \emph{square gradient term} is one of
the simplest ways to model the surface tension explicitly. Such terms have
been used all the way back to pioneering works of van der Waals on
one-component systems~\cite{Waals1979JSP} and by Cahn and Hilliard on
two-component systems~\cite{Cahn1958JCP}. Particularly relevant for
the hPF method is its recent implementation in \emph{hPF-Brownian
  dynamics} to model polymer-air
interfaces~\cite{Sgouros2018Macro}. Here, we reformulate the interaction energy for hPF simulations, 
also including anisotropic square gradient terms, to allow for an appropriate representation of 
the equation of state, making it possible to simulate constant pressure conditions.

\section{$NPT$ ensemble Hybrid Particle-Field}\label{sec:NPT-ensemle}
\subsection{hPF Hamiltonian}
We consider a system formed by $N$ molecules subject to the following
Hamiltonian:
\begin{equation} \label{eq:hamil}
  H=\sum\limits_{m=1}^{N}H_0(\{\v r,\dot{\v r}\}_m)+W[\{\phi(\v r),\gv\nabla\phi\}].
\end{equation}
$H_0$ is the single-particle Hamiltonian for the $m$-th molecule:
\begin{equation}
    H_0=T(\{\dot{\v r}\}_m)+U_0(\{\v r\}_m),
\end{equation}
where $T$ and $U_0$ are its kinetic and intramolecular potential
energies. In hPF models, intermolecular interactions are typically taken 
into account by the
interaction energy functional $W$, which is implicitly dependent on
the position of the particles through the set of number densities $\{\phi_k\}$, where the index $k$ indicates a particle type. Here
we introduce a new formulation of the energy functional, making it
also dependent on density gradients $\{\gv\nabla\phi_k\}$. We separate
the interaction energy into two terms:
\begin{equation}
  W[\{\phi_k(\v r),\gv\nabla\phi_k\}]=W_0[\{\phi_k\}]+W_1[\{\gv\nabla\phi_k\}].
\end{equation}
\paragraph{W$_0$: Flory-Huggins mixing entropy and compressibility}
The original formulation for hPF under $NVT$ conditions employed the
following interaction energy functional
~\cite{Milano2009JCP,Milano2010JCP}:
\begin{equation}\label{eq:energy}
W_0[\{\phi_k(\v{r})\}] = \frac{1}{2\phi_0}\int \text{d}\v{r}
\left( \sum_{k\ell} \tilde \chi_{k\ell} \phi_k(\v{r})
\phi_\ell(\v{r}) + \frac{1}{\kappa} \left( \sum_\ell \phi_\ell(\v{r})
- \phi_0 \right)^2 \right),
\end{equation}
where 
$\phi_0$
is the average total number density, $\tilde\chi_{k\ell}$ is the
Flory-Huggins coupling parameter between species $k$ and $\ell$,
and $\kappa$ controls the fluctuations of the local density. To
generalize this formulation to $NPT$ conditions, we propose the
following modified interaction energy:
\begin{equation}\label{eq:energy2}
W_0[\{\phi_k(\v{r})\}] = \frac{1}{2\rho_0}\int \text{d}\v{r} \left(
\sum_{k\ell} \tilde \chi_{k\ell} \phi_k(\v{r}) \phi_\ell(\v{r}) +
\frac{1}{\kappa} \left( \sum_\ell \phi_\ell(\v{r}) - a \right)^2
\right).
\end{equation}
Here $\rho_0=1/v_0$ is a constant related to the scale of coarse
graining, where $v_0$ is the molecular volume of the coarse grained
particles. $a$ is an independent parameter of the equation of state
with the dimension of a number density.
The corresponding external potential is given by:
\begin{equation} \label{eq:potential}
V_{0,k}(\v{r}) = \frac{\delta W_0[{\phi_k(\v{r})}]}{\delta
  \phi_k(\v{r})} = \frac{1}{\rho_0} \left( \sum_{\ell}
\tilde\chi_{k\ell} \phi_{\ell}(\v{r}) + \frac{1}{\kappa} \left(
\sum_\ell \phi_\ell(\v{r})- a\right)\right).
\end{equation}
We emphasize that because the parameter $a$ gives a constant contribution the potential $V_{0,k}$, it does
not affect the forces acting on the particles. We also note
that in the case of $\rho_0 = \phi_0 = a$, this new potential becomes strictly the same as the one used in the $NVT$ formulation.

\paragraph{$W_1$: Square gradient interactions}
To model interfaces we introduce a square gradient term to the
interaction energy~\cite{Sgouros2018Macro,Onuki2007PRE} dependent on
multiple species:
\begin{equation}
  W_1[\gv\nabla\phi]=\frac 1{2\rho_0}\sum_{k,\ell}\int\text d \v r~  K_{k\ell}  \gv\nabla\phi_k(\v r)\cdot\gv\nabla\phi_\ell(\v r),
\end{equation}
where $K_{k\ell}$ is a coupling constant between the gradients of
species $k$ and $\ell$. The corresponding external potential is given
by (see \siref{SI:V1-deriv}):
\begin{align}
  V_{1,k}(\v r) =-\sum_{\ell=1}\frac{ K_{k\ell}}{\rho_0}\nabla^2\phi_\ell(\v r).
\end{align}
\subsection{Calculation of the pressure in hPF} 
We calculate the pressure using a derivation similar to the one used by
H{\"u}nenberger for the reciprocal space part of Ewald
summation~\cite{Hunenberger2002JCP}. The pressure inside a simulation
box with side lengths $L_\mu$ and volume $V$ is given by:
\begin{equation}
  P_{\mu} = \frac{2T_\mu-\text{Vir}_\mu(\{\v r,\v L\})}{V}
\end{equation}
where $T_\mu$ denotes a Cartesian component of kinetic energy and
\begin{equation}
  \text{Vir}_\mu=L_\mu\frac{\partial U\name{tot}}{\partial L_\mu}
\end{equation}
is obtained directly from the potential energy of the system
$U\name{tot}$, defined as:
\begin{equation}
U\name{tot} = \sum_{m=1}^N U_0(\{\v r\}_m) + W_0[\{\phi\}] + W_1[\{\gv\nabla\phi\}]. 
\end{equation}
The bonded interactions ($U_0(\{\v r\}_m)$) contribute to the virial
term as in ordinary molecular dynamics. The interaction energy contributions to the
pressure are computed as:
\begin{equation}
  P_{0,\mu}=-\frac {L_\mu}V\frac{\partial W_0[\{\phi\}]}{\partial L_\mu},\quad
  P_{1,\mu}=-\frac {L_\mu}V\frac{\partial W_1[\{\gv\nabla\phi\}]}{\partial L_\mu},
\end{equation}
corresponding to (see \siref{appendix:virial-interaction} for their
derivation):
\begin{subequations}\label{eq:hpf:press}
\begin{align}
  P_{0,\mu}&=\frac 1V \int\text d \v r~\frac{1}{\rho_0}\left(\frac 12
  \sum_{k\ell}\tilde\chi_{k\ell}\phi_k(\v r)\phi_\ell(\v
  r)+\frac{1}{2\kappa}\left(\left(\sum_\ell \phi_\ell(\v{r})\right)^2-a^2\right)\right),\\ P_{1,\mu}& = \frac{1}V \int\text d \v
  r~\sum_{k\ell}\frac{K_{k\ell}}{\rho_0}\left(\frac 12\gv \nabla
  \phi_k(\v r)\cdot\gv \nabla \phi_\ell(\v r)+\nabla_\mu \phi_k(\v
  r)\nabla_\mu \phi_\ell(\v r) \right).\label{eq:p0p1}
\end{align}
\end{subequations} 
The total pressure in a direction $\mu$ is thus given by:
\begin{equation}\label{eq:press}
  P_\mu=\frac{2T_\mu}{V}+\frac{1}V\sum_i\left[-\pd{U_0(\v r_{i})}{
    r_{i,\mu}}\cdot r_{i,\mu}\right] +P_{0,\mu}+P_{1,\mu}
\end{equation}
Here we note the following: {\it (i)} Although $a$ gives no contributions to the
force, it gives rise to a nonzero pressure. This gives added
flexibility to control the isotropic pressure, similarly to the constant term in the
stiffened gas equation of state~\cite{Le2016PF}. {\it (ii)} The
contribution of $W_0$ to the pressure is isotropic, while the
contribution of $W_1$ is not. {\it (iii)} The local pressure density
(the integrand in \eqref{eq:p0p1}) does not contain a Laplace term as
reported in refs.~\cite{Onuki2007PRE,Sgouros2018Macro}. However, as
shown in \siref{appendix:virial-interaction}, the expressions are
equivalent.

\section{Computational details}
\subsection{hPF-MD simulations}
The model  described in the previous
section was implemented into hPF-molecular dynamics software OCCAM~\cite{Zhao2012JCP}. This enables the possibility of sample configurations of the molecular system governed by our new
hPF Hamiltonian \eqref{eq:hamil} following directly the evolution of the corresponding equations of motion.
The
forces on the $i$-th particle of type $k$ due to $W_0$ and $W_1$ are
computed from the gradients of the external potentials $V_{0,k}(\v r)$
and $V_{1,k}(\v r)$:
\begin{equation}
  \v F_{0,i}=-\gv\nabla V_{0,k}(\v r_i),\quad \v F_{1,i}=-\gv\nabla V_{1,k}(\v r_i).
\end{equation}
by a particle mesh approach~\cite{Nicolia2011JCTC}. First, particles are distributed onto a
different Cartesian grid for each species $k$ by linear interpolation
to the nearest vertices (cloud-in-cell).  Derivatives are computed on
a staggered grid by finite differences. Finally, the derivatives are
interpolated back onto the particles giving the forces. As
shown in ref.~\cite{Nicolia2011JCTC}, the external potentials are slow
variables, and can be updated with good approximation at intervals of
up to $\sim$100 steps~\cite{Nicolia2011JCTC}, yielding efficient
parallelization~\cite{Zhao2012JCP}.

\subsection{Computation of square gradient forces}
The computation of the external potential due to the square gradient
term $W_1$ involves computing the Laplacian of the densities. To
obtain a rotational invariant estimate without the appearance of
spurious oscillations, we employ a spectral
approach~\cite{canuto2006spectral} filtering out fast oscillations of the derivatives in
Fourier space. The filtering is done by convolution:
\begin{equation}
  \tilde \phi_k(\v r)=\int\text d \v u~ \phi_k(\v r-\v u)H(\v u).
\end{equation}
The tilde symbol denotes filtered densities by the applied filter
$H$. The corresponding external potential is given by:
\begin{align}
  V_{1,k}(\v r) =-\sum^M_{\ell=1}\frac{ K_{k\ell}}{\rho_0}\int \text d \v y~H(\v r-\v y)\nabla^2\tilde\phi_\ell(\v y).
\end{align}
which takes the following simple expression in Fourier space (see
\siref{SI:V1-deriv} for its derivation):
\begin{equation}
 \hat V_{1,k}(\v q) =\sum^M_{\ell=1}\frac{ K_{k\ell}}{\rho_0}q^2H^2(\v q)\hat\phi_\ell(\v q).
\end{equation}
Real space values are computed by backwards Fast Fourier Transform
routines (FFT). For consistency, the gradients in $P_1$ are also filtered
with the same filter $H$. Details on the filtering algorithm are
provided in \siref{SI:filter}.

\subsection{Barostat}\label{appendix:barostat}
We employ the Berendsen barostat~\cite{Berendsen1984JCP} with
isotropic coupling for isotropic systems, and semiisotropic coupling
for lipid bilayers. The efficiency of the hPF-MD approach is dependent
on having \emph{i)} little communication among processors and
\emph{ii)} avoiding heavy calculation (typically involving the grid)
between density updates.  Similarly to the multi-time-step approach
used in the GPU version of \emph{Tinker-OpenMM}~\cite{Harger2019RIC},
we average contributions from bonded terms and keep volume and field
contributions constant between density update steps.

\subsection{Simulation details}
We tested our model on a set of homogeneous and inhomogeneous
systems. Details on the composition of each individual system, as well
as information about other simulation parameters are given in
\siref{SI:sim-details}. The data that support the findings of this study are available from the corresponding author upon reasonable request.
\section{Results and discussion}
\subsection{Homogeneous system: Water}
\begin{figure}[!htb]
  \centering
  \includegraphics{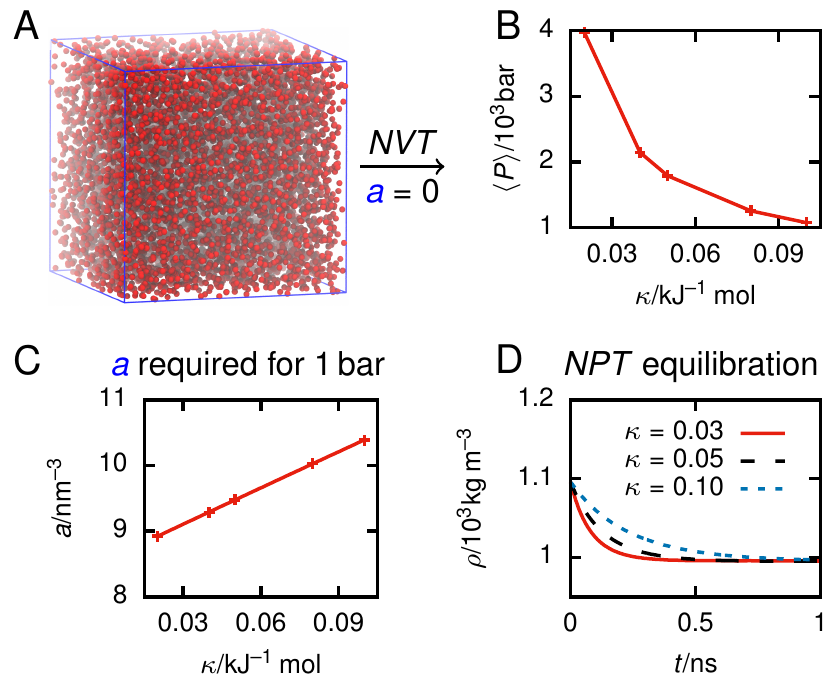}
  \caption{Parameterization of water for $NPT$ simulation. (A)
    Simulation box of water. (B) Pressure as function of $\kappa$ for
    $a=0$. (C) Calibration of $a$ to obtain an internal pressure of \SI{1}{bar} as function of
    $\kappa$. The least square fit line is $a(\kappa)=8.54+18.61\kappa$. (D)
    Equilibration of a overly dense liquid by $NPT$ simulation.}
  \label{fig:water}
\end{figure}
Within the hPF model, the representation of a homogeneous phase requires the consideration of
the interaction energy $W_0$ only. Furthermore, considering a
single-component system, the forces are only  dependent on its compressibility term.  In \figref{fig:water} we report the
parameterization of pure liquid water employing the commonly used explicit bead
model in hPF-MD where four water molecules are mapped into a single
body~\cite{Nicolia2011JCTC} (\figref{fig:water}A). Such a mapping
implies a molecular volume per bead $v_0=\SI{0.120}{nm^3}$,
thereby $\rho_0=1/v_0=\SI{8.33}{nm^{-3}}$.  In \figref{fig:water}B,
the pressure under $NVT$ conditions is plotted as function of $\kappa$
for a system with a density of \SI{995}{kg.m^{-3}} and temperature of \SI{300}{K}, using
$a=\SI{0}{nm^{-3}}$. The combination of the two positive definite
kinetic energy and compressibility terms produce an average internal pressure $\left\langle P\right\rangle\gg$~\SI{1}{bar}. From \eqref{eq:press} and \eqref{eq:hpf:press}, it
is possible to predict the value of $a$:
\begin{equation}\label{eq:a-by-p}
  a=\sqrt{(\left<P\right>_{a=0}-P_0)/(2\kappa\rho_0)},
\end{equation}
that would yield an equilibrium value $\langle P\rangle_a= P_{0}$, where $P_0$ is any target pressure of choice. Inserting $P_0=\SI{1}{bar}$ and values of $\left<P\right>_{a=0}$ into \eqref{eq:a-by-p}, we find a parameterization of $a$ as function of $\kappa$ (\figref{fig:water}C) which yields a pressure of \SI{1}{bar} for a density of \SI{995}{kg.m^{-3}} at \SI{300}{K}.  The parameterization of $a(\kappa)$ is fitted well by a linear regression line. Using this regression line we find, for three commonly used values of $\kappa={0.03},~0.05,~\SI{0.10}{kJ^{-1}mol}$,  $a=9.10,~9.47,~\SI{10.40}{nm^{-3}}$ respectively. Having parameterized $a$, we can now  simulate the water model under $NPT$ conditions. \figref{fig:water}D reports the time evolution of total mass density $\rho$ under $NPT$ conditions with barostat pressure of \SI{1}{bar} and temperature of \SI{300}{K} for a water system with an initial density of \SI{1100}{kg.m^{-3}}. For the three values of $\kappa$ the density equilibrates to the correct density. In the next systems, if not otherwise stated, we adopt $\kappa=\SI{0.05}{kJ^{-1}mol}$ and $a=\SI{9.47}{nm^{-3}}$.

\subsection{Binary mixture}
\paragraph{Phase separation}
\begin{figure}[!htb]
  \centering
  \includegraphics{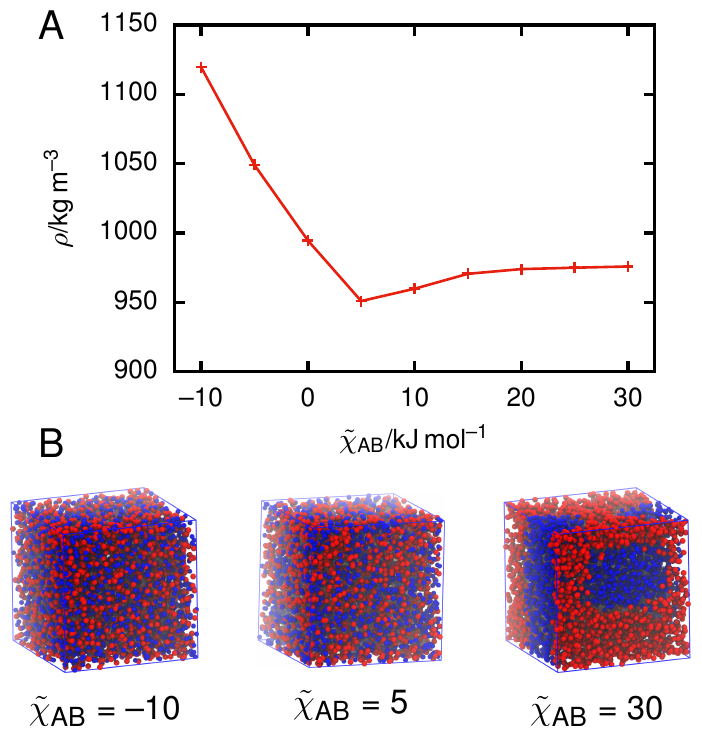}
  \caption{(A) Density of a binary mixture of two ideal fluids A, B of equal density as function of $\tilde\chi\name{AB}$. (B) Representative snapshots from hPF-MD after equilibration, highlighting the phase behavior for different  $\tilde\chi\name{AB}$ (values of $\tilde\chi\name{AB}$ are in kJ~mol$^{-1}$).}\label{fig:binary-mixture-chi}
\end{figure}
We consider a toy binary mixture between two ideal fluids. The two
components differ only by $\tilde\chi\name{AB}$ in the
potential energy term $W_0$. In \figref{fig:binary-mixture-chi} we
survey the state of this mixture by plotting its total density
as a function of $\tilde\chi\name{AB}$ at \SI{1}{bar}. 
The total density of
the mixture exhibits a strong excess volume effect, where the density
of the mixture is different from its components. For negative values
of $\tilde\chi\name{AB}$, mixing of the two fluids is favourable, and the
density increases. For positive values of $\tilde\chi\name{AB}$ the density
is lower and stabilizes to a constant value for high values of $\tilde\chi\name{AB}$. The stabilization can be
interpreted from the snapshot at $\tilde\chi\name{AB}=\SI{30}{kJ.mol^{-1}}$
as the formation of a sharp interface between the two phases. The
abrupt change in the first derivative of the total density at about
$\tilde\chi\name{AB}=\SI{5}{kJ.mol^{-1}}$ signals a phase transition. This
is further evidenced by the snapshots showing a transition from
miscible to  phase-separated fluids before and after
$\tilde\chi\name{AB}=\SI{5}{kJ.mol^{-1}}$.

\paragraph{Ideal water/oil droplet}
\begin{figure}[!htb]
  \centering
  \includegraphics[width=3.4in]{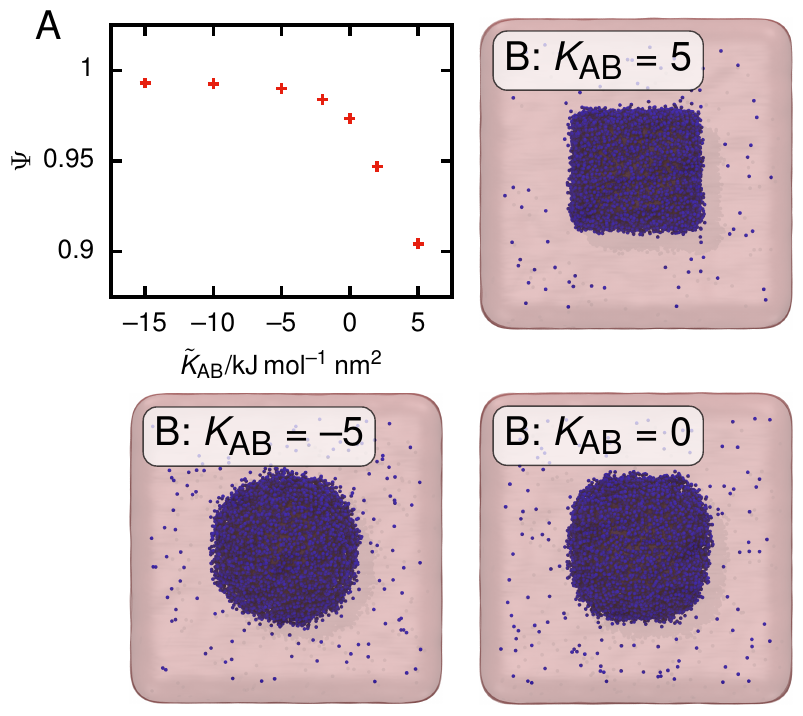}
  \caption{ hPF-MD simulations of a droplet of an immiscible liquid A in liquid B ($\tilde\chi\name{AB}=\SI{20}{kJ.mol^{-1}}$), using 
    different values of $K\name{AB}$. (A) Sphericity of the droplet as function of $K\name{AB}$. (B) Snapshots of the dropble after equilibrated by hPF-MD, liquid B enveloping the droplet is represented by a transparent red surface. Values of $K\name{AB}$ are in kJ~mol$^{-1}$~nm$^2$.}\label{fig:binary-mixture-kst}
\end{figure}
While the $\tilde \chi\name{AB}$ term in $W_0$ controls the partitioning and the level of phase
separation between the two liquids, the interaction energy $W_1$ is necessary for modelling interfacial properties, and in particular surface tension. In the case of a binary system, $W_1$ requires
the definition of only one parameter $K\name{AB}$ to
control the surface interaction between the two
phases. We survey how $K\name{AB}$ affects interfaces by simulating an ideal oil droplet (particle type A) in water (particle type B, constituting 90\% of the particles in the simulation)  and by computing its sphericity for different values of $K\name{AB}$ (\figref{fig:binary-mixture-kst}A). The sphericity $\Psi$ is defined by the equation~\cite{Wadell1935JG}:
\begin{equation}
    \Psi \equiv\frac{\pi^{1/3} 6V^{2/3}}{A},
\end{equation}
where $A$ and $V$ are the surface area and the volume of the droplet. For very negative values of $K\name{AB}$, we find a sphericity close to 1, corresponding to almost a perfect sphere (snapshot in \figref{fig:binary-mixture-kst}B). This is consistent with a sphere having the lowest possible surface for a given volume. By increasing $K\name{AB}$, we lower the interfacial energy. This allows for larger surface areas of the droplet, and thus the appearance of other shapes than a sphere. In our simulations, for $K_{AB}=\SI{0}{kJ.mol^{-1}.nm^{2}}$, we found a configuration in between sphere and cube and for $K\name{AB}=\SI{5}{kJ.mol^{-1}.nm^{2}}$ we observed a configuration very close to a cube (snapshots in \figref{fig:binary-mixture-kst}B). We note that the formation of a cube is likely affected by the orientation of the grid used to to calculate particle-field forces, which has been reported to produce cube shaped vesicles\cite{Sevink2017soft}.  

\subsection{Effect of square gradient term on lipid bilayers}
\begin{figure}[!htb]
  \centering \includegraphics{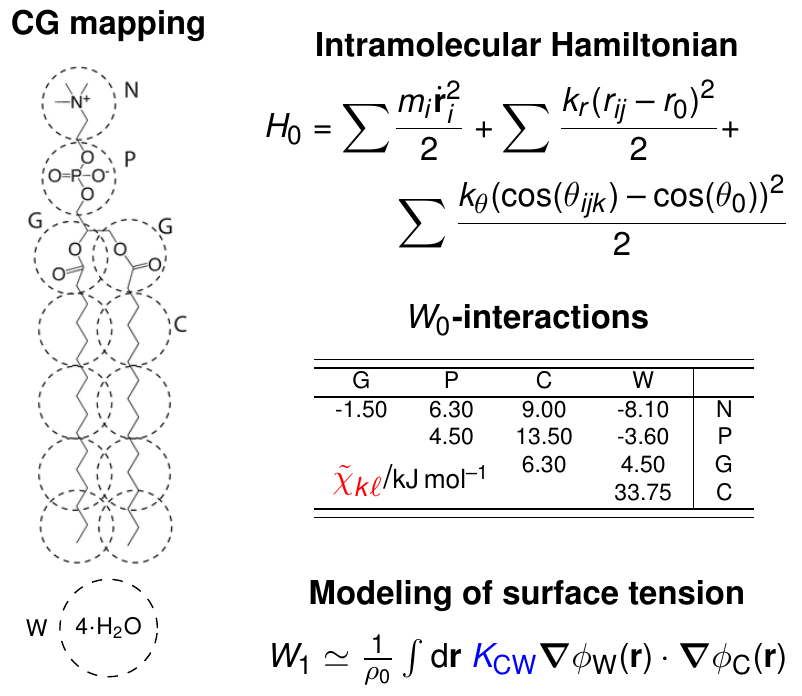}
  \caption{{\it Left:} CG model used for hPF-MD simulations of DPPC lipid bilayers in water. {\it Right:} functional form of the intra-molecular potential for DPPC, and particle-field interaction terms used in the simulations.}\label{fig:lipid-model}
\end{figure}
To test the feasibility of the proposed approach to models with specific molecular features, we investigate a realistic model of a
dipalmitoylphosphatidylcholine (DPPC) lipid bilayer in water, employing a
molecular CG representation, and the corresponding $\tilde\chi$
interaction energy matrix present in the
literature~\cite{Nicolia2011JCTC}, as summarized in
\figref{fig:lipid-model}. Here, we add to the preexisting model
the square gradient interaction limited to only one $K_{m\ell}$ term between the hydrophobic lipid tail (C) and the water (W) beads
($K\name{CW}$), disregarding all other terms.

\paragraph{flat lipid bilayers -- Surface area}
\begin{figure}[!htb]
  \centering
  \includegraphics[width=3.4in]{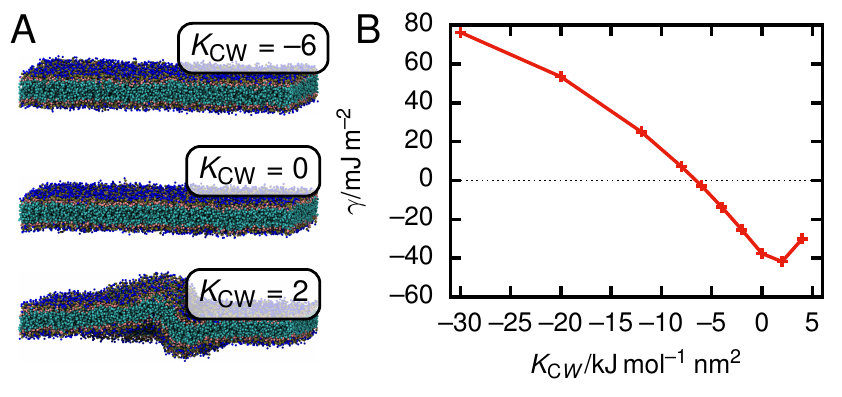}
  \caption{$NVT$ simulations of DPPC lipid bilayer.   (A)
    Snapshots of equilibrated membranes using different $K\name{CW}$ values. (B) Surface
    tension of DPPC lipid bilayer as a function of $K\name{CW}$.}
  \label{fig:NVT-DPPC}
\end{figure}
$NVT$ simulations of lipid bilayers in periodic boundary conditions
impose an arbitrary effective area per
lipid $A$, defined as:
\begin{equation}
    A =  \frac{2L_x L_y}{N},
\end{equation}
where $L_x, L_y$ are the edges of the simulation box in the $x,y$
directions spanning lipid bilayer, and $N$ is the number of assembled
lipids. In this case, $NVT$ simulations allow for a controlled study
of the effects of $K\name{CW}$ on the morphology of the
system. \figref{fig:NVT-DPPC}A reports equilibrated conformations for
different values of $K\name{CW}$. As we start from preoptimized
$\tilde\chi$ values to reproduce flat bilayers at
$K\name{CW}=\SI{0}{kJ.mol^{-1}.nm^{2}}$, a negative value of
$K\name{CW}=\SI{-6}{kJ.mol^{-1}.nm^{2}}$ does not produce strong
structural modifications. On the contrary,
$K\name{CW}=\SI{2}{kJ.mol^{-1}.nm^2}$ induces an abrupt change in the
bilayer with the formation of visible bump within the first
\SI{190}{ns} of simulations. This deformation is consistent with the
fact that positive values of $K\name{CW}$ promote the expansion of the
interface area.

This trend can be quantified by computing the surface tension $\gamma$
of the membrane, \figref{fig:NVT-DPPC}B, which can be calculated from:
\begin{equation}
  \gamma=\frac 12\int\text d z~\left(P_{\text N}(z)-P_{\text L}(z)\right).
\end{equation}
Here, $P_{\text N}(z)$ and $P_{\text L}(z)$ are the values of the pressure
in the normal and lateral directions of the membrane plane,
respectively. The $1/2$ factor takes into account the presence of two
interfaces. A negative value of $\gamma$ in the absence of the square
gradient interaction energy indicates that the area per lipid is not
at equilibrium, and the system would tend to expand laterally if let
free to relax. $K\name{CW}\sim\SI{-6}{kJ.mol^{-1}.nm^{2}}$ balances the
two pressures, and should predict an equilibrium area per lipid at
$NPT$ conditions equal to the initial target value.

\begin{table}[!ht]
  \caption{Predicted area per lipid compared against literature data from experiment and
    simulations.}\label{tab:area}
  \centering {\footnotesize
  \begin{tabular}{lcccccccc}
    \hline\hline
    Source&\multicolumn{2}{c}{DMPC}&\multicolumn{2}{c}{DPPC}&\multicolumn{2}{c}{DSPC}&\multicolumn{2}{c}{DOPC}\\
    &$A/\si{nm^2}$&$T/\si{K}$&$A/\si{nm^2}$&$T/\si{K}$&$A/\si{nm^2}$&$T/\si{K}$&$A/\si{nm^2}$&$T/\si{K}$\\
    \hline
    Nagle~\cite{Nagle2000BBA}
    &0.596&323&0.630 &323 &-&-&0.725&303\\
    Waheed~\cite{Waheed2009BJ} sim&0.625&303&0.644&323&-&-&-&-\\
    Waheed~\cite{Waheed2009BJ} exp&0.606&303&0.630&323&-&-&-&-\\
    Levine~\cite{Levine2014JACS} &-&- & 0.629&323&-&-&0.689&298\\
    Petrache~\cite{Petrache2000BJ}&0.600&303&0.633&323&0.66&338&-&-\\
    &0.654&323&0.671&338&-&-&-&-\\
    \hline
    hPF-MD&0.61&323&0.64&325 & 0.66&338 & 0.70&303\\
    \hline\hline
  \end{tabular}
}
\end{table} 
We thus simulated a $\SI{20}{nm}\times\SI{100}{nm}\times\SI{100}{nm}$
large DPPC/water system at $NPT$ condition employing the same
$a=\SI{9.59}{nm^{-3}}$ determined for pure water at \SI{325}{K} (see
\siref{SI:a-para}) and using $K\name{CW}=\SI{-6}{kJ.mol^{-1}.nm^2}$.
After an initial relaxation, the DPPC bilayer reaches an equilibrium
configuration characterized by a well defined area per lipid of
$\SI{0.64}{nm^2}$ (Table~\ref{tab:area}). This value is in excellent
agreement with what has been previously reported in the
literature\cite{Nagle2000BBA,Waheed2009BJ,Petrache2000BJ}.  

The transferability of K$\name{CW}$ was tested on three other lipids, namely: dimyristoylphosphatidylcholine (DMPC), distearoylphosphatidylcholine (DSPC) and  dioleoylphosphatidylcholine (DOPC), which differ from DPPC only for the length of the carbon tail while retaining the same chemical structure of the polar head.
As for DPPC, we used the W$_0$ parameters from ref.~\cite{Nicolia2011JCTC}; hPF-MD $NPT$ simulations ran using the  
same $a = 9.59$~nm$^{-3}$, $K\name{CW}=-6$~kJ~mol$^{-1}$. Remarkably, these parameter produce in all cases  excellent agreement with literature data, as presented in Table~\ref{tab:area}, indicating indeed a high level of transferability among chemically similar moieties. 

\paragraph{Flat lipid bilayers -- membrane structure}
\begin{figure*}[!ht]
\centering
\includegraphics{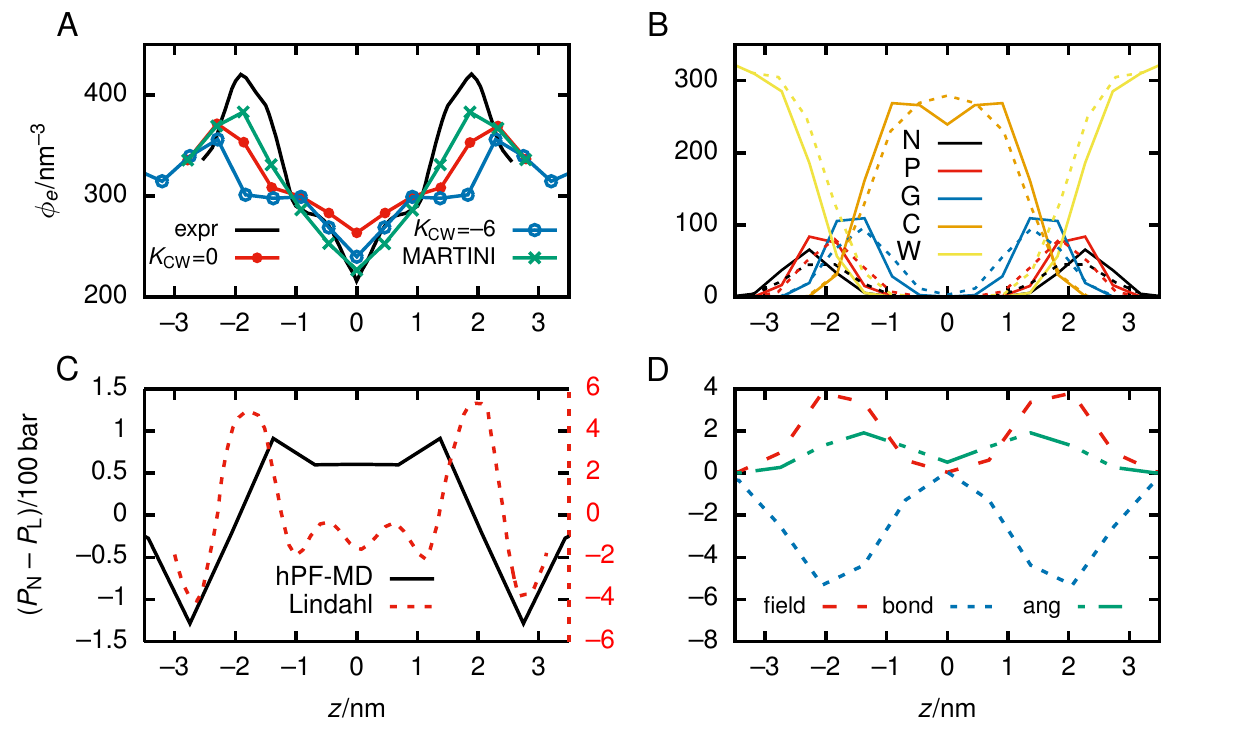}
\caption{Density and pressure profiles for DPPC. (A) Comparison between the experimental electron density profilea (black line),  hPF-MD using $K\name{CW}$=0~kJ~mol$^{-1}$nm$^2$ (red line), $K\name{CW}=\SI{-6}{kJ.mol^{-1}.nm^2}$ (blue line)
and reference CG MARTINI model~\cite{Nicolia2011JCTC}. (B) Density profiles for the different
  bead types. Continuous lines are for hPF-MD data using
  $K\name{CW}=\SI{-6}{kJ.mol^{-1}.nm^2}$,  dashed lines are by De Nicola et al.~\cite{Nicolia2011JCTC} at
  $NVT$ conditions without a square gradient term. (C) Difference
  between normal and lateral components of the pressure for tensionless hPF-MD
  simulations compared against all-atom pressure profiles by Lindahl
  and Edholm~\cite{Lindahl2000JCP}. Two different scales are used for
  the $y$-axis for the two models. (D) Contributions to the pressure difference shown in (C) from
  density field, bonded, and angular
  interactions.}\label{fig:dppc-profiles}
\end{figure*}
We survey the effect of $K\name{CW}$ on the equilibrium structure of
the bilayer by computing the electron density profiles along the
membrane normal axis for a small DPPC lipid bilayer (See
\siref{SI:syst-setups} for a detailed system description).
\figref{fig:dppc-profiles}A reports a comparison between the density
profiles from experiment, a CG simulation using the MARTINI force
field\cite{Marrink2007JCPB}, reference hPF simulations under $NVT$
without the square gradient term, and hPF in $NPT$ with
$K\name{CW}=\SI{-6}{kJ.mol^{-1}.nm^2}$. All the profiles exhibit the
peaks at the head and low electron density in the middle of the
bilayer. The profile obtained with
$K\name{CW}=\SI{-6}{kJ.mol^{-1}.nm^2}$ has better agreement with
experiment and MARTINI in the middle of the bilayer, showing instead
some excessive elongation in the position of the polar head
beads. This trend is
corroborated by drawing the individual bead contributions to the
density profile, as shown in
\figref{fig:dppc-profiles}B. Such opposite trends are not entirely surprising, keeping in mind that
the square gradient term has been applied to the carbon tails only,
while the polar head have not been corrected by any surface tension
contribution. The current results suggest that by an appropriate
calibration of the whole $K_{\ell m}$ matrix, the square gradient term can improve significantly the
agreement between hPF and its underlying CG model. Interestingly, the peaks of the
individual beads appear sharper, indicating a more regular bilayer compared
to hPF $NVT$ simulations.

We also computed local pressure profile, using the values of $P_0$ and
$P_1$ at the vertices of the mesh, and Kirkwood-Irving assignment of
bonded virials.\cite {Sonne2005JCP} \figref{fig:dppc-profiles}C,D
report the local difference between normal and lateral pressure
computed with hPF-MD and all-atom by Lindahl,\cite{Lindahl2000JCP}
and the contributions by the different terms of the Hamiltonian. While
the magnitude of the all-atom pressure is about four times that of
hPF-MD, we nevertheless identify the appearance of three key qualitative features: the
presence of a negative peak between water and the lipid heads, a
positive peak in between heads and tails, and a rather flat region in
the tail part. Unlike all-atom simulations, the hPF profile has an positive sign to the pressure difference at the carbon tails,
indicating compressed carbon tails in the normal direction. We note
that this may be part an artifact of the coarseness of the mesh, which
can cause a spill out of pressure into the middle part of the
membrane, as well as by the absence of a square gradient terms between
water and glycerol or polar head beads. The local pressure
contributions from $P_0$ and $P_1$ were computed at the vertices of
the mesh. This is at one hand rigorous as it avoids subtleties related
to local pressure assignment, however higher resolution would be
advantageous for computing properties from the local pressure
profiles. A natural route for achieving higher resolution assignment
of pressure would be to follow the procedure proposed in
ref.~\cite{Sega2016JCTC,Sonne2005JCP} by Harasima
assignment.\cite{Harasima1958ACP}

 Our simulations present a very flat bilayer without the detection of
 significant undulations on the tensionless surface of the
 bilayer. The stiffness of the bilayer can be quantified by computing
 the \emph{area compressibility} as in ref.~\cite{Marrink2001JPCB}.

For the DPPC lipid bilayer, in our case, we obtain
$K_A=\SI{22000}{mN.m^{-1}}$. This is about two orders of magnitude larger than
what has been reported in the literature~\cite{Levine2014JACS}
($\sim\SI{200}{mN.m^{-1}}$) confirming that the present setup produces an
excessively rigid system.  Although it is well known that the
Berendsen barostat is not suited for studying fluctuations of the
membrane~\cite{Wong2016BBA}, we stress here, in these preliminary test implementation, that we have only
considered carbon water interactions for the square gradient
term. Moreover, this term was naively added to $\tilde\chi$ parameters that
were preoptimized to reproduce accurate density profiles in the
absence of an explicit surface tension term, with consequent possible
double counting of the repulsion between the water and the hydrophobic
tails.  Overall, the discrepancy on the fluctuation of the DPPC
bilayer together with the qualitative but not quantitative agreement
on the lateral pressure profiles indicate that $NPT$ simulation of
realistic systems require a global parameterization of both the $\tilde\chi$
and $K_{m\ell}$ matrices, while a simple addition of the second term
to the first may not be sufficient to obtain quantitatively accurate data.

\section{Conclusion and outlook}
We presented a reformulation of the hPF interaction energy suitable
for constant pressure simulations using both isotropic and anisotropic coupling. 
First, we
modified the commonly used interaction energy by introducing an
equation of state parameter $a$. By design, this adjustment conserves
the dynamics of the old formulation. Second, we introduced a square
gradient term to the interaction energy to model interfacial
phenomena. Particle-field
contributions to the pressure were derived by considering change in free energy
upon change in simulation box lengths. The equation of state parameter
$a$ enters as an added constant to the pressure. The square gradient
contributes to nonisotropic pressure, thereby allowing for direct
modeling of surface tension. Our approach  was implemented into
the OCCAM code, where the dynamics of system governed by the hPF
Hamiltonian was sampled by MD and pressure was coupled to the Berendsen
barostat.

Testing on simple single particle fluids demonstrated how by tuning
$a$ we can reproduce the densities at ambient conditions, also showing
how the $\tilde\chi$-term can be used to modulate variations in the partial
molar volume in liquid mixtures. We also verified that the square
gradient term can be tuned to control the shape of liquid droplets.

Finally, we tested the effect of the new hPF Hamiltonian on a
realistic model of a phospholipid bilayer previously proposed in the
literature.  Interestingly, the square gradient term is not only
important, but mandatory for achieving an area per lipid within the
experimental range under $NPT$ conditions, as well as a qualitatively
reasonable lateral pressure profile. Interestingly, we also found that the same parameterization of $a$ and $K\name{CW}$ is transferable to other
lipids of similar chemical composition.

Remarkably, the application of only one square gradient contribution
between the carbon tails and water was sufficient to obtain a
qualitatively correct physical behaviour of the lipids as well as some
impressive improvement of some of their key structural features like
average area per lipid, or lateral pressure profiles. Nonetheless, we
found that such correction produced inconsistent variations in the
lateral density profiles, and a too stiff bilayer, indicating that a
consistent recalibration of the $\tilde\chi$ parameters as well as the use
of the full $K_{m\ell}$ matrix is necessary for quantitative agreement
between hPF and other higher resolution models as well as the
experiment.

Accessing constant pressure conditions significantly expands the applicability of hPF simulations.
For example, it is now possible predict density changes in bulk systems, or to represent 
surface phenomena. The future challenge is in the calibration of appropriate square gradient force constant matrices, possibily through combined global parameterizations with the bulk enery terms, aiming for quantitatively accurate description of interfaces. 

\section{Acknowledgments}
The authors would like to acknowledge Morten Ledum for help with
generating initial bilayer structures.
\section{Funding}
Authors acknowledge the support of the Norwegian Research
Council through the CoE Hylleraas Centre for Quantum
Molecular Sciences (Grant No. 262695) and the Norwegian
Supercomputing Program (NOTUR) (Grant No. NN4654K). MC acknowledges funding by the
Deutsche Forschungsgemeinschaft (DFG) within the project B5 of the TRR 146 (project number
233630050).
HBK received funding from the European Union Horizon
2020 research and innovation program under the Marie
Sk{\l}odowska-Curie Grant Agreement HYPERBIO - No 704491.

\appendix
\section*{Supporting information}
In this supporting information we provide the details on
\emph{derivations}, \emph{computational procedures}, \emph{simulation
  setups} and \emph{parameterizations}, that are needed for
reproducing the results obtained in this manuscript.

\subsection{Derivations}
All derivations involving square gradient term are performed with
filtered densities $\tilde \phi$.

\subsubsection{External potential $V_{1,k}$}\label{SI:V1-deriv}
We compute the external potential by using the functional derivative
chain rule twice:
\begin{equation}\label{eq:V1-deriv-1}
  V_{1,k}(\v r) = \frac{\delta W_1[\gv \nabla \tilde \phi]}{\delta \phi_k(\v r)}= \int\text d \v x\text d \v y~\frac{\delta W_1}{\delta\gv \nabla\tilde\phi_k(\v x)}\frac{\delta\gv \nabla\tilde\phi_k(\v x)}{\delta\tilde\phi_k(\v y)}\frac{\delta\tilde\phi_k(\v y)}{\delta\phi_k(\v r)}.
\end{equation}
The terms in the integrand of \eqref{eq:V1-deriv-1} are given by:
  \begin{equation}
    \frac{\delta W_1}{\delta\gv \nabla\tilde\phi_k(\v x)}=\sum^M_{\ell=1} \frac{K_{k\ell}}{\rho_0}\gv\nabla\tilde\phi_\ell(\v x),\quad
    \frac{\delta\gv \nabla\tilde\phi_k(\v x)}{\delta\tilde\phi_k(\v y)}=\gv\nabla_{\v x} \delta(\v x -\v y),\quad   \frac{\delta\tilde\phi_k(\v y)}{\delta\phi_k(\v r)}=H(\v r-\v y).
  \end{equation}\label{eq:V1-integrand}
Inserting \eqref{eq:V1-integrand} into \eqref{eq:V1-deriv-1} yields:
\begin{align*}
  V_{1,k}(\v r) &= \int\text d \v x\text d \v y~\sum^M_{\ell=1} \frac{K_{k\ell}}{\rho_0}\gv\nabla\tilde\phi_\ell(\v x)\gv\nabla_{\v x} \delta(\v x -\v y)H(\v r-\v y)\\
  &=\int \text d \v y~H(\v r-\v y)\int\text d \v x~\sum^M_{\ell=1} \frac{K_{k\ell}}{\rho_0}\gv\nabla\tilde\phi_\ell(\v x)\gv\nabla_{\v x} \delta(\v x -\v y)\\
  &= -\int \text d \v y~H(\v r-\v y)\int\text d \v x~\sum^M_{\ell=1} \frac{K_{k\ell}}{\rho_0}\nabla^2\tilde\phi_\ell(\v x) \delta(\v x -\v y)\\
  &=-\sum^M_{\ell=1} \frac{K_{k\ell}}{\rho_0}\int \text d \v y~H(\v r-\v y)\nabla^2\tilde\phi_\ell(\v y).
\end{align*}
Without filter ($H(\v r-\v y)=\delta(\v r-\v y)$) we obtain:
\begin{equation}
    V_{1,k}(\v r)=-\sum^M_{\ell=1} \frac{K_{k\ell}}{\rho_0}\nabla^2\tilde\phi_\ell(\v r)
\end{equation}
\subsubsection{Virial terms from interaction energies}\label{appendix:virial-interaction}
The pressure is given by the Viral~\cite{Hunenberger2002JCP}:
\begin{equation}
  P_{\mu} = \frac{2K_\mu-\text{Vir}_\mu(\{\v r,\v L\})}{V}
\end{equation}
where
\begin{equation}
  \text{Vir}_\mu=L_\mu\frac{\partial U}{\partial L_\mu}.
\end{equation}
Focusing on nonbonded terms and starting with $W_0$, we denote the
integrand by interaction energy density $w(\{\phi(\v r)\})$, we
compute derivative with respect to box size:
\begin{align}
  \pd{W_0}{L_\mu} & = \int \left(\pd{\text d \v
    r}{L_\mu}~w_0(\phi(\v r))+\text d \v r ~\pd{w_0(\phi(\v
    r))}{L_\mu}\right)\\
  &= \int \left(\frac{\text d \v
    r}{L_\mu}~w_0(\phi(\v r))-\text d \v r ~\pd{w_0(\phi(\v
    r))}{\phi}\frac{\phi(\v r)}{L_\mu}\right)
\end{align}
giving:
\begin{equation}
  \text{Vir}_{0,\mu}=\int \text d \v r \left(w_0(\{\phi(\v r)\})- \sum_{\ell=1}^M\pd{w_0(\{\phi(\v r)\})}{\phi_\ell(\v r)}\phi_\ell(\v r)\right).
\end{equation}
Inserting the energy density:
\begin{equation}
   \text{Vir}_{0,\mu}=- \int\text d \v r~\frac{1}{\rho_0}\left(\frac 12
  \sum_{k\ell}\tilde\chi_{k\ell}\phi_k(\v r)\phi_\ell(\v
  r)+\frac{1}{2\kappa}\left(\phi(\v r)^2-a^2\right)\right)
\end{equation}

We compute interface virial of $W_1$ by:
\begin{align}
  \pd{W_1}{L_\mu}&=\int \left(\pd{\text d \v r}{L_\mu} ~w([\gv \nabla \tilde \phi(\v r)])+\text d \v r ~\pd{w([\gv \nabla \tilde \phi(\v r)])}{L_\mu}\right)\\
  &=\int \left(\frac{\text d \v r}{L_\mu} ~w([\gv \nabla \tilde \phi(\v r)])+\text d \v r ~\pd{w([\gv \nabla \tilde \phi(\v r)])}{\nabla_\nu \tilde \phi(\v r)}\pd{ \nabla_\nu \tilde \phi(\v r)}{L_\mu}\right).
\end{align}
The last partial derivative is given by:
\begin{align}
  \pd{}{L_\mu} \left(\nabla_\nu \tilde \phi(\v r)\right) &= \left(\pd{}{L_\mu}\nabla_\nu\right)\tilde \phi(\v r)+\nabla_\nu \pd{\tilde \phi(\v r)}{L_\mu}\\
  & = -\frac{\delta_{\mu\nu}}{L_\mu}\nabla_\nu\tilde \phi(\v r)+\nabla_\nu\frac{\partial\tilde \phi(\v r)}{\partial L_\mu}.
\end{align}
The partial derivative filtered density with respect to box lengths
depends on the filter. This dependence is easily understood by using
the Fourier transform:
\begin{align}
  \pd{\tilde \phi(\v r)}{L_\mu}=&\pd{}{L_\mu}\int\text d\v q~ \tilde\phi(\v
  q)e^{-i\v q\v r}= \pd{}{L_\mu}\int\text d\v q~ \hat\phi(\v q)\hat H(\v
  k)e^{-i\v q\v r}\\
  =&\int\left(\pd{\text d\v q}{L_\mu}\hat\phi(\v q)\hat H(\v
  k)e^{-i\v q\v r}+\text d\v q\pd{\hat\phi(\v q)}{L_\mu}\hat H(\v
  k)e^{-i\v q\v r}\right.+\\
  &\left.\text d\v q\hat\phi(\v q)\hat H(\v
  k)\pd{e^{-i\v q\v r}}{L_\mu}+\text d\v q\hat\phi(\v q)\pd{\hat H(\v
  k)}{L_\mu}e^{-i\v q\v r}\right).
\end{align}
The contributions from the three first terms are given by:
\begin{equation}
  \pd{\text d \v q}{L_\mu}=-\frac{\text d \v q}{L_\mu},\quad
  \pd{\hat\phi(\v q)}{L_\mu}=0,\quad \pd{e^{-i\v q\v r}}{L_\mu}=0.
\end{equation}
Whether the last term contributes, depends on the specifics of the
filter. The filter we employ is of the form:
\begin{equation}
  H(\v q)\equiv H(\v q \cdot \v l),
\end{equation}
where $\v l$ is the cell size, which means the filter is independent of
box size, and thus:
\begin{equation}
  \pd{\hat H(\v  q)}{L_\mu}=0.
\end{equation}
Therefore we have
\begin{equation}
  \pd{\tilde \phi(\v r)}{L_\mu} = -\frac{\tilde \phi(\v r)}{L_\mu}
\end{equation}
which results in the following expression:
\begin{equation}
  \text{Vir}_{1,\mu} = \int\text d \v r~\left(w_1(\{\gv \nabla\tilde \phi(\v r)\})-\sum_{\ell=1}^M\left(\pd{\tilde \phi_\ell(\v r)}{L_\mu}+\gv \nabla\tilde \phi_\ell(\v r) \right)\pd{w_1([\gv \nabla \tilde \phi(\v r)])}{\gv \nabla \tilde \phi_\ell(\v r)}    \right).
\end{equation}
For the square gradient term, the virial term is given by:
\begin{equation}\label{eq:virial-SI}
\text{Vir}_{1,\mu} = -\int\text d \v r~\sum_{k\ell}\frac{K_{k\ell}}{\rho_0}\left(\frac 12\gv \nabla\tilde \phi_k(\v r)\gv \nabla\tilde \phi_\ell(\v r)+\nabla_\mu\tilde \phi_k(\v r)\nabla_\mu\tilde \phi_\ell(\v r)  \right).
\end{equation}
\paragraph{Consistency with literature}
The pressure computed in~\cite{Onuki2007PRE,Sgouros2018Macro} corresponds to a virial of the form:
\begin{equation}\label{eq:onuki}
\text{Vir}_{1,\mu} = -\int\text d \v r~\sum_{k\ell}\frac{K_{k\ell}}{\rho_0}\left(
-\frac 12 \tilde \phi_k(\v r)\nabla^2\tilde \phi_\ell(\v r))-\frac 12\gv \nabla\tilde \phi_k(\v r)\gv \nabla\tilde \phi_\ell(\v r)\right.\\\left.+
\nabla_\mu\tilde \phi_k(\v r)\nabla_\mu\tilde \phi_\ell(\v r)  
\right).
\end{equation}
We rewrite \eqref{eq:onuki} as:
\begin{equation}\label{eq:onuki2}
\text{Vir}_{1,\mu} = -\int\text d \v r~\sum_{k\ell}\frac{K_{k\ell}}{\rho_0}\left(
-\gv\nabla\left(\tilde \phi_k(\v r)\gv \nabla\tilde \phi_\ell(\v r)\right)+\frac 12\gv \nabla\tilde \phi_k(\v r)\gv \nabla\tilde \phi_\ell(\v r)\right.\\\left.+
\nabla_\mu\tilde \phi_k(\v r)\nabla_\mu\tilde \phi_\ell(\v r)  
\right).
\end{equation}
where we have used:
\begin{equation}
    \phi\nabla^2\phi=\gv\nabla(\phi\gv\nabla\phi)-\gv\nabla\phi\gv\nabla\phi.
\end{equation}
Finally, for periodic boxes, the integral of the gradient sums to zero giving:
\begin{equation}\label{eq:onuki3}
\text{Vir}_{1,\mu} = -\int\text d \v r~\sum_{k\ell}\frac{K_{k\ell}}{\rho_0}\left(
\frac 12\gv \nabla\tilde \phi_k(\v r)\gv \nabla\tilde \phi_\ell(\v r)+
\nabla_\mu\tilde \phi_k(\v r)\nabla_\mu\tilde \phi_\ell(\v r)  
\right),
\end{equation}
which is the same as \eqref{eq:virial-SI}.

\subsection{Computational details}

\subsubsection{Computation of Laplace term}\label{SI:filter}
The forces from the gradient term involves a gradient of the Laplace
operator. As hPF-MD uses a coarse grid with distribution of particles
to only neighbouring grid points, special numerical techniques are
required to avoid amplification of unphysical high frequency modes for
higher order derivatives. We introduce the following regularized
density variable:
\begin{equation}
  \tilde \phi_k(\v r)=\int\text d \v u~ \phi_k(\v r-\v u)H(\v u),
\end{equation}
where $H(\v u)$ is a normalized distribution often referred to as a
kernel, window function or transfer function. Using the spectral
method the derivative is obtained to arbitrary order through:
\begin{equation}
 \gv \nabla^n \tilde \phi_k(\v r)=\text{FFT}^{-1}\left[(i\v q)^n \hat\phi_k(\v q)\hat H(\v q)e^{i\v q\cdot\v r}\right],
\end{equation}
where $\hat{}$ denotes variable in Fourier space. In the literature
many transfer functions are reported, some more commonly used are
raised cosine and Gaussian filter. Our main interest lies in
computation of second order derivative, therefore we use the following
specialized second-order filter:
\begin{equation}
  \hat H(\v q, \v l) = \frac{1}{\sqrt{1+(\left|\v q\cdot\v l\right|)^4}}. 
\end{equation}
\subsection{Simulation details}\label{SI:sim-details}
Here we provide details on all the systems simulated. Unless otherwise
specified for a specific system, parameters in \siref{SI:sim-para} are
employed.
\subsubsection{Simulation procedures and parameters}\label{SI:sim-para}
Constant temperature simulations are achieved by the
\emph{Andersen thermostat}~\cite{Andersen1980JCP} with a collision
frequency of \SI{7}{ps^{-1}} and a coupling time of \SI{0.1}{ps}.  For
$NPT$ simulations, pressure is kept constant by the Berendsen barostat
with a compressibility parameter set to \SI{4.5E-5}{bar} with a
coupling time of \SI{12}{ps}. Equations of motion are integrated using
the \emph{velocity Verlet algorithm}~\cite{William1982JCP} with
time step \SI{0.03}{ps}. The densities used for computing the
particle-field forces are updated every \SI{3}{ps}. For all
simulations $\rho_0=\SI{8.33}{nm^{-3}}$. The number of cells used is chosen such that their lengths
are $\sim\SI{0.67}{nm}$.
\subsubsection{System setups}\label{SI:syst-setups}
\paragraph{Water simulations}
The pressure graph in presented in \figref{fig:water}, is obtained by
simulating a cubic box of size
$\SI{15}{nm}\times\SI{15}{nm}\times\SI{15}{nm}$ containing 28113
beads under $NVT$ conditions at \SI{300}{K}. The system was first equilibrated for \SI{15}{ns} and data was
then gathered for \SI{15}{ns} gathering pressure every
\SI{0.15}{ns}. Next the  $NPT$ equilibration of density was performed on
the same box size, but with 30915 beads. In this specific simulation,
a coupling time constant for the barostat of \SI{12}{ps} was used.
\paragraph{Binary mixtures}
The binary mixture results presented in
\figref{fig:binary-mixture-chi}, a box of
$\SI{25}{nm}\times\SI{25}{nm}\times\SI{25}{nm}$ containing 130156
beads with a 50\%/50\% mixture of type A and B. The system was first equilibrated for
\SI{15}{ns} and the data was then gathered for \SI{15}{ns} gathering
pressure at every \SI{0.15}{ns}. For the compressibility term a
$\kappa=\SI{0.05}{kJ^{-1}mol}$ was used. For snapshots of the droplets
presented in \figref{fig:binary-mixture-kst} are obtained with the same
box only starting from 10\%/90\% mixture with
$\tilde\chi\name{AB}=\SI{20}{kJ.mol^{-1}}$. 
\paragraph{Lipid bilayers}
\subparagraph{$NVT$}
As starting configuration for the \figref{fig:NVT-DPPC}, a highly
undulating membrane solvated in water in a box of
$\SI{40}{nm}\times\SI{40}{nm}\times\SI{20}{nm}$ was used. The membrane is
composed out of 5000 lipids with 12 beads each and 206400 water beads
(corresponding with this mapping to 825600 water molecules). This
specific membrane is kept at \SI{325}{K}.  The membrane was simulated
for a total time of \SI{190}{ns}, and data was gather from \SI{30}{ns}
every \SI{0.3}{ns}.
\subparagraph{$NPT$}
The starting configuration was prepared by the \emph{insane-code}
~\cite{Wassenaar2015JCTC}, with an initial box of
$\SI{100}{nm}\times\SI{100}{nm}\times\SI{20}{nm}$ containing 33282
lipids and 1238345 water beads (corresponding to 4953380 water
molecules). A coupling time of \SI{12}{ps} was used
for the barostat. The system was first equilibrated for \SI{15}{ns},
then the cells for the density grids are updated to fit the new box
and a second simulation lasting \SI{150}{ns} is performed. The profiles presented in \figref{fig:dppc-profiles} were obtained for a system of 528 DPPC lipids solvated with 24000 water beads (960000 water molecules) 
with an initial equilibration of \SI{15}{ns} followed by a simulation of \SI{90}{ns} of data gathering every \SI{75}{ps}.

\subsection{Parameterization of temperature dependence of {\it a} for water}\label{SI:a-para}
\begin{figure}[!htb]
  \centering \includegraphics{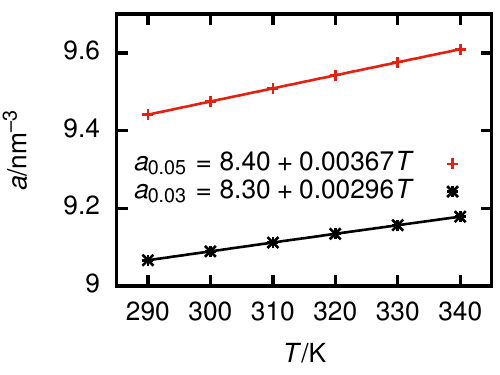}
  \caption{Required $a$ to reach \SI{1}{bar} for
    $\kappa=\SI{0.03}{kJ^{-1}.mol}$ and $\SI{0.05}{kJ^{-1}.mol}$ for
    different temperatures.}\label{fig:a-temp-water}
\end{figure}
The parameterization of $a$ is obtained by considering a box of
water beads with density \SI{995}{kg.m^3} under
$NVT$ conditions. Keeping $a=0$, a pressure $P$ as a function of the
parameters is found. Using \eqref{eq:a-by-p}, the required $a$ to get the correct density at
\SI{1}{bar} is obtained. \figref{fig:a-temp-water} shows for two
commonly used compressibility values the required $a$ as function of
temperature.
\bibliography{pressure}
\end{document}